\documentstyle[12pt,epsf]{article}

\newcommand{\be}{\begin{equation}}
\newcommand{\ee}{\end{equation}}
\newcommand{\bea}{\begin{eqnarray}}
\newcommand{\eea}{\end{eqnarray}}
\newcommand{\ba}{\begin{array}}
\newcommand{\ea}{\end{array}}
\newcommand{\tLambda}{\tilde{\Lambda}}
\newcommand{\tm}{\tilde{m}}

\newcommand{\journal}[4]{{\rm #1} {\bf #2} (19#3) #4}
\newcommand{\NP}{\journal{Nucl. Phys.}}

\makeatletter
  
  \@addtoreset{equation}{section}
\makeatother

\begin{document}

%
%
\begin{titlepage}
\begin{flushright}
OU-HET 290 \\
hep-th/9803123 \\
March 1998
\end{flushright}
\bigskip
\bigskip

\begin{center}
{\Large \bf Orientifold 4-plane in brane configurations\\ and \\$N=4\;  USp(2N_c)$ and $SO(N_c)$ theory
}

\bigskip
\bigskip
\bigskip
Takashi Yokono\footnote{e-mail address: yokono@funpth.phys.sci.osaka-u.ac.jp}\\
\bigskip
{\small \it
Department of Physics,\\
Graduate School of Science, Osaka University,\\
Toyonaka, Osaka 560, JAPAN
}
\end{center}
\bigskip
\bigskip
\bigskip
\bigskip
\bigskip

\begin{abstract}
We consider brane configurations in elliptic models which represent softly broken $N=4 \; USp(2 N_c)$ and $SO(N_c)$ theory. We generalize the notion of the O4 plane, so that it is compatible with the symmetry in the covering space of the elliptic models. By using this notion of the O4 plane, we find the curve for softly broken $N=4 \; USp(2 N_c)$ and that for $SO(N_c)$ theory as infinite series expansions. For the $USp$ case, we can present the expansion as a polynomial. 
\end{abstract}

\end{titlepage}

%
%
\section{Introduction}
Supersymmetric Yang-Mills theory are interpreted as low-energy worldvolume theory on branes \cite{pol}. By considering brane configurations, some properties of SYM such as dualities are explained in terms of branes \cite{hw, egkrs, gk}. Conversely,  we may know behaviors of branes by studying corresponding supersymmetric field theory.

In particular N=2 super QCD is given by worldvolume theory on D4 branes ending on NS 5-branes in type IIA string theory. In M-theory, NS5 and D4 branes are the same object, namely M5-brane, which represents ${\bf R}^{1,3}\otimes\Sigma$ \cite{witten}. Here ${\Sigma}$ is complex Riemann surface which is described by Seiberg-Witten curves \cite{sw}. The N=2 Higgs branch is also studied in terms of M-theory \cite{hoo, noyy}. The softly broken $N=4\quad SU(N_c)$ model has been also considered in \cite{witten}. This model is given by compactifying the direction along D4 branes which begin from and end on single NS5 brane (elliptic models). This space is twisted, which means that even if we go around the compactified direction, we do not come back to the same point but to the shifted point along the NS5 brane. The mass of the matter in the adjoint representation is given by the magnitude of the shift.

The curves for $N=2\quad SO(N_c)$ and that for $USp(2 N_c)$ theory have been derived \cite{lll, bsty} by introducing an orientifold 4-plane in brane configurations. How to introduce orientifold 4-planes in elliptic models with twist is,however, remains as a question till now.

In this paper we consider brane configurations with the `O4 plane' in elliptic models representing softly broken $N=4 \; USp(2 N_c)$ and $SO(N_c)$ theory. In the conventional definition of O4 planes, the mass to the matter is not permitted. This is because ${\bf Z}_2 $ projection of the O4 plane and the shift which corresponds to the mass are not compatible globally. So we need to generalize O4 plane projection. To do this, we first of all consider the O4 plane in each fundamental region in the covering space of elliptic models. Curves for softly broken $N=4 \; USp(2N_c) $, $SO(2N_c)$ and $SO(2N_c+1)$ are given by  infinite series expansions. They are given by (\ref{usp}), (\ref{so}) and (\ref{so2}) in the text respectively. These equations are consistent with the decoupling limits. In the $USp$ case, we can read off the symmetry which the `O4 plane' induces in the compact space since we can represent the curve as the polynomial. This is given by (\ref{uspp}). Although we have not represented $SO$ curve as polynomials in this paper, we may see that the curve has the symmetry as well. In this `O4 plane' case, the mass of the matter is given by the shift along NS5 brane as in the $SU$ case.

\section{Elliptic Models and the interpretation as the covering space}

In this section we review \cite{witten} how $N=4\quad SU(N_c)$ theory is realized in the brane configuration. This model is equivalent to what is provided by considering the integrable system \cite{dw,im,dp}. In particular  D'Hoker and Phong \cite{dp} give the curve as the expansion in $q$ , where
\be
q=e^{2\pi i \tau}.
\ee
Here $\tau$ is complex gauge coupling constant,
\be
\tau=\frac{\theta}{2\pi} + \frac{4 \pi i}{g^2}.
\ee
This expansion of the curve in q may be interpreted as the brane configuration in the covering space.

\begin{figure}[t]
\epsfysize=5cm \centerline{\epsfbox{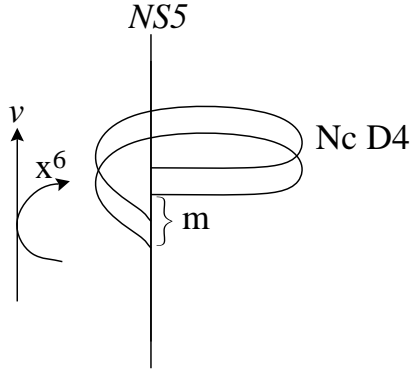}}
\caption{\small
The elliptic model.
}
\label{elliptic}
\end{figure}

First of all, we explain the space in which the M5 brane is embedded. We consider only single NS5 brane and $N_c$ D4 branes. In IIA picture, the NS5 brane has worldvolumes in the $x^0,x^1,x^2,x^3,x^4,x^5$ directions and its location is specified by the $x^6,x^7,x^8,x^9$. D4 branes have worldvolumes in the $x^0,x^1,x^2,x^3,x^6$ directions and their locations are specified by the  $x^4,x^5,x^7,x^8,x^9$. These branes are merely an M5 brane in M-theory picture. Nonetheless we call D4 or NS5 depending upon whether M5 is winding in the $x^{10}$ direction or not. The $x^{10}$ direction is compactified on $S^1$ of radius $R$. We introduce holomorphic coordinates such that 
\bea
v&=&x^4 + i x^5\\
s&=&\frac{x^6 + i x^{10}}{R}.
\eea
We compactify the $x^6$ direction on a circle of radius $L$. Let us consider the following configuration: the NS5 brane is local on this circle and $N_c$ D4 branes begin from and end on this NS5 brane and extend along this circle. The $v-s$ space is ${\bf C}\times {\bf E}$ locally, where $v \in {\bf C},\: s\in {\bf E}$. Here {\bf E} is a genus one Riemann surface with an arbitrary complex structure. However we should consider a ${\bf C}$ bundle over ${\bf E}$,
\bea
x^6 &\to & x^6 + 2\pi L ,\nonumber \\
x^{10} &\to & x^{10} - \theta R,\label{xmtf} \\
v &\to & v+m.\nonumber
\eea
We call this bundle $X_m$. If $m=0$ then we may take $X_m = {\bf C}\times {\bf E}$ globally. The parameter $m$ corresponds to the mass of the matter in the adjoint representation. See Fig \ref{elliptic}.
If we set $\tau = \frac{\theta}{2\pi} + i\frac{L}{R}$, therefore $\frac{4\pi}{g^2} = \frac{L}{R}$, the above first two lines in (\ref{xmtf}) are expressed as
\be
s \to  s - 2 \pi i \tau .
\ee
The softly broken $N=4\; SU(N_c)$ curve is interpreted as an $N_c$-fold cover of {\bf E}, the $N_c$ branches being the positions of the D4 branes in {\bf C}. The NS5 brane appears as a simple pole in $v$. This description is equivalent to the integrable system \cite{witten}.

We begin by saying that the explicit shape of the M5 brane is given by the  integrable system. If we set double period $(2w_1,2w_2)=(1,\tau)$ \footnote{This choice does not change physical charges} the curve is given by \cite{dp}
\bea
0=f(k,z)&=&\det (k- \frac{m}{\beta}h_1(z)-L(z)) \nonumber \\
&=&\sum_{n=0}^{N_c}h_n(z)Q_{N_c-n}(k), \label{cm}
\eea
where 
\bea
k &=& v - \frac{m}{2}, \\
z &=& \frac{s}{\beta} ,\\
\beta &=& -2 \pi i ,\\
h_n(z) &=& \frac{1}{\theta_1(z|\tau )} \frac{\partial^n}{\partial z^n}\theta_1(z|\tau ) ,\label{hn} \\
Q_{N_c - n}(k) &=& \frac{(-m)^n}{n!\beta^n} \frac{\partial^n}{\partial k^n}H(k) ,\label{qh} \\
H(k)&=& \prod_{a=1}^{N_c}(k - k_{a} ).
\eea
Here $L(z)$ is the gauge transformed Lax operator in the elliptic Calogero-Moser system \cite{kr}. The $k_a$ correspond to the classical locations of the D4 branes in $v$-plane and at the same time are vevs of the adjoint scalar in $N=2$ vector multiplet. The curve is embedded in $X_m$ since the curve (\ref{cm}) is invariant under the transformation (\ref{xmtf}).\footnote{Notice that $h_1(z+\tau) = h_1(z)+\beta$}

Using the series expansion for $\theta_1(z|\tau )$, the curve is expanded as the infinite series in powers of $q=e^{2 \pi i \tau}$ up to overall phase factors which do not depend on $n$,
\be
\sum_{n=-\infty}^{\infty}(-)^n q^{\frac{1}{2}n(n-1)}e^{n s}H(v- n m)=0 .\label{dhsu}
\ee
\begin{figure}[t]
\epsfysize=6cm \centerline{\epsfbox{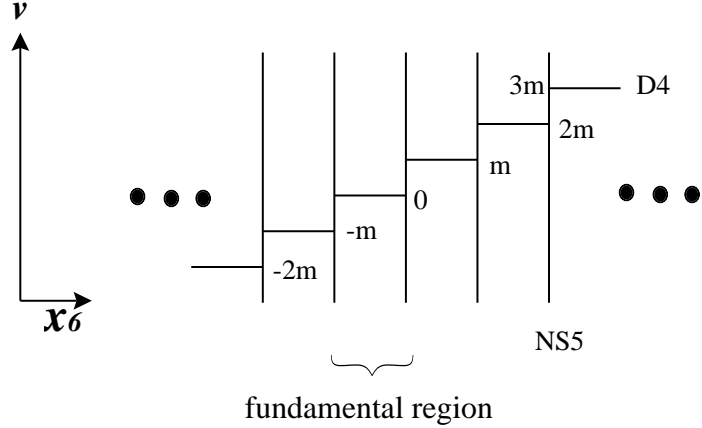}}
\caption{\small
The brane configuration in the covering space.
}
\label{covering}
\end{figure}
This expression is faithfully represented by the brane configuration in the covering space of $X_m$ (Fig \ref{covering}). The reason is that in noncompact $x^6$ space the curve for $N=2 \; \bigotimes^{n}SU(N_c)$ gauge group with bifundamental matters theory is given by the following form \cite{witten},
\be
y^{n+1} + B_{n-[\frac{n+1}{2}]}(v) y^n + \ldots + B_{0}(v)y^{[\frac{n+1}{2}]}+ \ldots + B_{1-[\frac{n+1}{2}]}(v) y + 1 =0.
\ee
Here $y=e^s$ and $B_i(v)$ are some polynomials of degree $N_c$ in $v$. Dividing the curve by $y^{[\frac{n+1}{2}]}$, we obtain
\be
y^{n+1-[\frac{n+1}{2}]} + \ldots + B_{1}(v) y + B_{0}(v) + B_{-1} (v)\frac{1}{y} +\ldots + \frac{1}{y^{[\frac{n+1}{2}]}} =0.
\ee
If we set $B_{l}(v)=B(v-lm)$, D4 branes located between a pair consisting an NS5 brane and the next NS5 brane get shifted by $m$ in the $v$ direction when they move to the adjacent pair of NS5 branes. The mass of each bifundamental matter is $m$. Let $n\to \infty $ and D4 branes located between each pair of NS5 branes be the same. Therefore $\bigotimes^{n}SU(N_c)$ is reduced to single $SU(N_c)$ and bifundamental matters become a hypermultiplet in the adjoint representation with mass $m$ (plus a neutral singlet). This curve is equivalent to (\ref{dhsu}) and this configuration represents the covering space of the softly broken $N=4\; SU(N_c)$ in the elliptic model.

\section{Orientifold 4-plane in elliptic models}

Let us introduce an O4 plane to the elliptic model by considering the covering space. We treat an O4 plane as a nondynamical D4 brane which carries appropriate  R-R charges and which operates ${\bf Z_2}$ space projection \cite{lll,bsty}. Being nondynamical means that the O4 plane does not give rise to new moduli. The ${\bf Z_2}$ space projection is $(v,x^7,x^8,x^9) \to (-v,-x^7,-x^8,-x^9)$. The difference in worldsheet parity projection is the difference in R-R charges. If the R-R charge of the O4 plane is +1 then the gauge group is $USp$, and if -1 then $SO$.

It is clear that we cannot put an O4 plane as the projection into the elliptic model when the mass does not vanish. If we give a ($v\leftrightarrow -v)$ mirror symmetry to the configuration as a constraint, there is a conflict between this symmetry and another symmetry (\ref{xmtf}). The reason is clear in the covering space (Fig \ref{covering}). Mirrors of D4 branes around $v=nm$ are produced around $v=-nm$ in $n_{\mbox{th}}$ fundamental region and around $v=(n+1)m$ and $v=-(n+1)m$ in the next one. We must, however, have D4 branes around $v=(n+1)m$ and $v=-(n-1)m$ in the next one according to (\ref{xmtf}).

\begin{figure}[t]
\epsfysize=6cm \centerline{\epsfbox{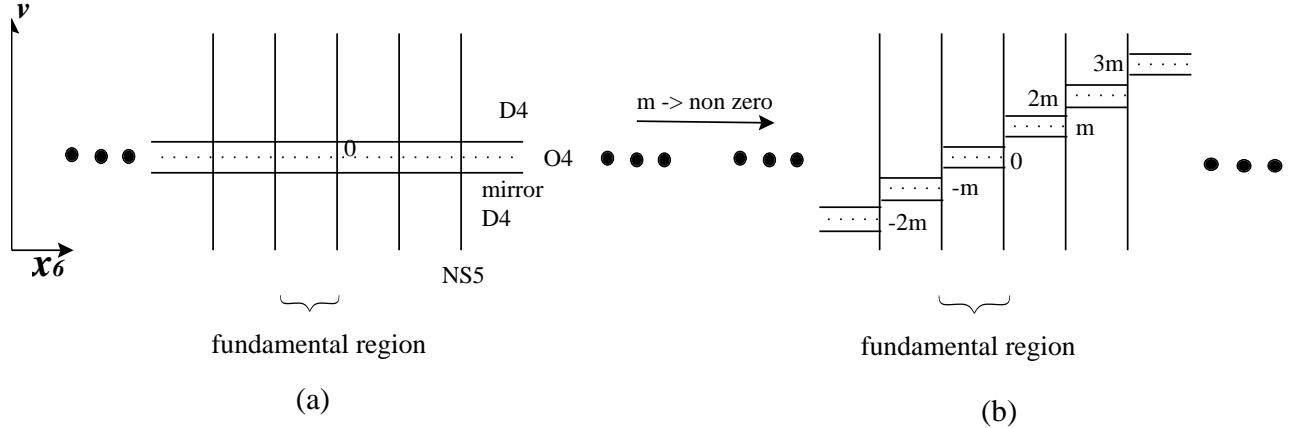}}
\caption{\small
When $m=0$ an O4 plane exits globally. Turning on $m$, D4 branes and the O4 plane move together.  
}\label{covering2}
\end{figure}

Hence we require the `O4 plane' projection as the O4 plane projection in each fundamental region,
\be
v-nm \to -(v-nm) \mbox{ for each $n$}.\label{O4}
\ee
This seems a natural generalization of the O4 plane since we have normal O4 plane when $m=0$. Turning on $m$, each fundamental region is shifted by $m$ along the $v$ direction. This time each O4 plane is shifted with other D4 branes together. See (Fig \ref{covering2}).
This configuration apparently corresponds to softly broken $N=4\; USp$ and $SO$ theory with the O4 plane whose R-R charge is +1 and -1 respectively. In what follows we read off curves from this configuration and check some consistency.

\subsection{$ N=4 \; USp(2 N_c)$ theory}

Let us consider the case that the gauge group is $USp(2 N_c)$. There are $N_c$ D4 branes, $N_c$ mirror D4 branes and an O4 plane whose R-R is +1 in each fundamental region. Let $k_a > 0,\quad a=1,\ldots ,N_c$ be positions of $N_c$ D4 branes in the $n=0$ fundamental region. The positions of mirror D4 branes are $-k_a$. Hence $H(v-nm)$ in (\ref{dhsu}) has $\prod_{a=1}^{N_c}\left( (v-nm)^2 - k_a^2\right)$ part for D4 branes. This polynomial is further deformed by the O4 plane. There are two effects of O4 plane in $H(v-nm)$ whose R-R charge is +1 between two NS5 branes. The one is to introduce $(v-nm)^2$ in $H(v-nm)$. The other is to introduce some constant shift which is independent of both $v$ and $s$. For example \cite{lll,bsty}, pure $N=2 \quad USp(2 N_c)$ case, the curve with an O4 plane whose position is $v=0$ is 
\be
y^2 + y\left( v^2\prod_{a=1}^{N_c}\left( v^2 - k_a^2\right)+2 \right) + 1 =0,
\ee
where $y=e^s$. The first $v^2$ is caused by the O4 plane with the following assumption \cite{ejs}: if the orientifold crosses a NS5 brane its charge changes sign. In our case, to be exact, the O4 plane does not cross, but the same term should be induced in each fundamental region. Hence we have $H(v-nm)=(v-nm)^2\prod_{a=1}^{N_c}\left( (v-nm)^2 - k_a^2\right) + a_n$, where $a_n=a_n(q,m,k_a)$. This is invariant under (\ref{O4}) for each $n$ as expected. The curve (\ref{dhsu}) is deformed to 
\be
\sum_{n=-\infty}^{\infty}(-)^n q^{\frac{1}{2}n(n-1)}e^{n s}\left( (v-nm)^2\prod_{a=1}^{N_c}\left( (v-nm)^2 - k_a^2\right) + a_n \right) =0. 
\ee
for softly broken $N=4\;  USp(2 N_c)$. Let us consider the term $a_n$. Notice that this curve should have all symmetries that the curve (\ref{dhsu}) has for arbitrary $N_c$ because this configuration with the O4 plane is seen as the special case of $N=4 \; SU(2 N_c +2)$ configuration.\footnote{Actually, $USp(2 N_c) \subset SU(2 N_c) \subset SU(2 N_c + 2)$} Since the curve must be invariant under (\ref{xmtf}), we have $a_n = a_{n+1}$, which means that $a_n(q,m,k_a)=a(q,m,k_a)$. The curve (\ref{dhsu}) is also invariant under both $(n,m,s)\to (-n,-m,-s+2\pi i \tau)$ and $(v,m,k_a) \to (-v,-m,-k_a)$ so that $a(q,m,k_a) = a(q,m^2,k_a^2)$. The O4 plane is nondynamical. When $m=0$, the effects except for ${\bf Z_2}$ projection of the O4 plane in each $n_{\mbox{th}}$ fundamental region will be canceled by the ones from  $n+1_{\mbox{th}}$ and $n-1_{\mbox{th}}$ fundamental regions. This means that $a(q,m^2,k_a^2)=m^2 \tilde{a}(q,m^2,k_a^2)$.  The mass dimensions of $(q,m,k_a,a)$ are $(0,1,1,2N_c +2)$, respectively. Of course $\tilde{a}$ must be invariant under Weyl transformation of $USp(2 N_c)$. These constraints are satisfied,
\be
a(q,m^2,k_a^2) = f(q)m^2 \prod_{a=1}^{N_c}(m^2-k_a^2),
\ee
where $f(q)$ is some function such that $f(q)\to q^{\frac{1}{2}}$ for $q \to 0$. This formula will be checked in the next subsection. We find softly broken $N=4 \quad USp(2 N_c)$ curve as the infinite series expansion
\be
\sum_{n=-\infty}^{\infty}(-)^n q^{\frac{1}{2}n(n-1)}e^{n s}\left( (v-nm)^2\prod_{a=1}^{N_c}\left( (v-nm)^2 - k_a^2\right) + f(q)m^2 \prod_{a=1}^{N_c}(m^2-k_a^2) \right) =0.
\label{usp} 
\ee

\subsection{Decoupling limits}

We check the consistency of the curve (\ref{usp}) by the diagram below.\\
\begin{tabular}{ccc}
$N=4 \quad USp(2N_c)$ & $\longrightarrow $ & $N=4 \quad SU(N_c)$ \\
 & (1) & \\
$\downarrow $(2) & & $\downarrow $ (4) \\
&(3) & \\
$N=2\quad USp(2 N_1)$ with  $N_2$ flavors & $\longrightarrow $& $N=2 \quad SU(r_1)$ with $r_2$ flavors\\
 &  &
\end{tabular}

The limit (4) is already studied by D'Hoker and Phong \cite{dp}. It is known that when $r_1 = r_2$ the flavor group is gauged so that we have $N=2\; SU(r_1)\times SU(r_2)$ theory with bifundamental matters. We will see the similar situation in the limit (2).

For the limit (2), the curve for $N=2\quad USp(2 N_1)$ with $N_2$ flavors is already known \cite{as}. Therefore we need further checks for the case that the flavor symmetry is gauged.

Let us check the limit (1). We take all D4 branes far away from the O4 plane. In this limit, $v_0=\frac{1}{N_c}\sum k_a \to \infty$, where $v_0$ is the center of D4 branes not including mirrors in $v$-plane. We also shift the origin, which induces $v\to v + v_0$ and $k_a \to k_a + v_0$. The limiting form of $H(v-nm)$ is \footnote{Notice that we ignore the immaterial factor `2' which comes from the shifts of coordinates in this paper. Here this factor appears as $(v+v_0-nm+k_a+v_0)\to 2v_0$.}
\bea
H(v-nm) &=& (v-nm)^2\prod_{a=1}^{N_c}\left( (v-nm)^2 - k_a^2 \right) + f(q)m^2 \prod_{a=1}^{N_c}(m^2-k_a^2)  \nonumber \\
&\to& v_0^{N_c+2} (\prod_{a=1}^{N_c}(v-nm-k_a) + f(q)m^2 v_0^{N_c -2}). \nonumber
\eea  
Since we do not require $\sum k_a =0$ yet, we can absorb the last term into the first term by some redefinition of $k_a$, and then we may take $\sum k_a = 0$ by a further constant shift in $v$. The overall factor $v_0^{N_c +2}$
does not depend on $n$, which we throw away. Hence the limiting form of the curve is
\be
\sum_{n=-\infty}^{\infty}(-)^n q^{\frac{1}{2}n(n-1)}e^{n s}\prod_{a=1}^{N_c}(v-nm-k_a) =0
\ee
This is the same as (\ref{dhsu}), the curve for $N=4 \quad SU(N_c)$.

For the limit (2), we classify $k_a$ into two groups such that $x_i$ and $y_j+m$ where $i=1,\ldots,N_1,\quad j=1,\ldots,N_2$ respectively. Here $N_1 + N_2 = N_c$. In this limit, $q\to 0 (\tau \to i \infty)$ and $m\to \infty$ while keeping
\be
\Lambda^{b_0} = q m^{b_0},
\ee
fixed. Here $\Lambda $ is constant and $b_0 = 4 N_1 -2 N_2 + 4$.
\bea
H(v-nm)&=&(v-nm)^2 \prod_{i=1}^{N_1}((v-nm)^2-x_i^2)\prod_{j=1}^{N_2}(v-nm-m-y_j)(v-nm+m+y_j) \nonumber \\
&&+f(q)m^2 \prod_{i=1}^{N_1}(m^2-x_i^2)\prod_{j=1}^{N_2}(m-m-y_j)(m+m+y_j).\nonumber
\eea
We have the leading large m and small q behavior of $H(v-nm)$ ,
\bea
n=0\quad H&\sim &m^{2N_2}v^2\prod_{i=1}^{N_1}(v^2-x_i^2) + q^{\frac{1}{2}}m^{2N_1 + N_2+2}\prod_{j=1}^{N_2}y_j \nonumber \\
&\to &m^{2N_2}(v^2A(v) + \Lambda^{\frac{b_0}{2}}\prod_{j=1}^{N_2}y_j) ,\\
n=\pm 1 \quad H&\sim & m^{2+2N_1 + N_2}(\prod_{j=1}^{N_2}(v\pm y_j) + q^{\frac{1}{2}}\prod_{j=1}^{N_2}y_j) \nonumber \\
&\to& m^{2+2N_1 + N_2}B_{\pm}(v) ,\\
n \ne 0,\pm 1 \quad H &\to & m^{2N_1 + 2N_2 +2}, 
\eea
where
\bea
A(v)&\equiv & \prod_{i=1}^{N_1}(v^2-x_i^2) ,\\
B_{\pm }(v)&\equiv & \prod_{j=1}^{N_2}(v\pm y_j).
\eea
The limiting form of the curve is 
\bea
0=&&m^{2N_2}(v^2 A(v) + \Lambda^{\frac{b_0}{2}}\prod_{j=1}^{N_2}y_j) \nonumber \\
&&- \omega m^{2+2 N_1 + N_2} B_+ (v) + \omega^2 q m^{2N_1 + 2N_2 +2} + O(n > 2) \nonumber \\
&& -\omega^{-1} q m^{2N_1 + N_2 +2} B_-(v) + \omega^{-2} q^3 m^{2N_1 + 2N_2 +2} + O(n<-2),
\eea
where $\omega=e^s$. Since we need at least one NS5 brane, we keep $y$ fixed as  $m\to \infty$,
\be
y=\omega\: m^{\frac{b_0}{2}} .
\ee
For finiteness, we assume $-6N_1 + 4N_2 -6\le 0$. We obtain the limiting form of the curve as $ m\to \infty$ and $ q\to 0$ ,
\be
(v^2 A(v)+ \Lambda^{\frac{b_0}{2}}\prod_{j=1}^{N_2}y_j) - y B_+(v) - \frac{\Lambda^{b_0}}{y}B_-(v)+ y^2\Lambda^{b_0}m^{-6N_1 + 4N_2 -6} -\frac{\Lambda^{3b_0}}{y^2}m^{-6N_1 + 4N_2 -6} =0.
\ee
When $-6N_1 + 4N_2 -6 < 0$, we get the following curve as $m\to \infty $,
\be
(v^2A(v)+ \Lambda^{\frac{b_0}{2}}\prod_{j=1}^{N_2}y_j) - y B_+(v) - \frac{\Lambda^{b_0}}{y}B_-(v) =0 .\nonumber
\ee
This is the curve for $N=2\quad USp(2N_1)$ with $N_2$ flavors \cite{as} as expected. To be more explicit, by $y \to \frac{-1}{B_+(v)}y$, we obtain 
\be
y^2 + (v^2 \prod_{i=1}^{N_1}(v^2-x_i^2) +  \Lambda^{\frac{b_0}{2}}\prod_{j=1}^{N_2}y_j) y + \Lambda^{b_0}\prod_{j=1}^{N_2}(v^2-y_j^2) =0.
\ee
In particular, we take $N_2=0$ which means pure $N=2\; USp(2N_1)$ theory. The constraint, $-6 N_1 +4\cdot 0 -6 <0$, is satisfied. This curve is valid in this limit as well.

When $-6N_1 + 4N_2 -6 = 0$, $(N_1,N_2)=(2k-1,3k),\;  k \in{\bf Z}_{>0}$. The flavor group is gauged as in the $SU$ case \cite{dp}:
\be
\Lambda^{b_0} y^4 - B_+(v) y^3 + (v^2 A(v) + \Lambda^{\frac{b_0}{2}}\prod_{j=1}^{N_2}y_j) y^2 - \Lambda^{b_0} B_-(v) y + \Lambda^{3b_0} =0,\label{uspsu}
\ee
where $b_0= 4N_1-2N_2 + 4=2k$. This is the curve for $N=2\; USp(2N_1)\times SU(N_2)$ with bifundamental matters. To check consistency, we try the limit (3).

We consider the decoupling limits such that 
\be
USp(2N_1)\times SU(N_2) \mbox{ with  bifundamental matters}\quad \to \quad SU(r_1) \mbox{ with } r_2 \mbox{ flavors} \nonumber
\ee
Here we require $r_2 \le r_1$ as in \cite{dp}. The flavor group should be gauged if and only if $r_1 = r_2$. We classify $x_i,\; y_j$ into two groups. Let $M$ be some constant. The one is that $(x_{i_1} - x_{i_2})/M \to 0$ and $(y_{j_1} - y_{j_2})/M \to 0$ as $M\to \infty$, if $i_1,\; i_2$ and $j_1,\; j_2$ belong to the same group respectively. The other is that $(x_{i_1} - x_{i_2})/M \to \pm 1$ and $(y_{j_1} - y_{j_2})/M \to \pm 1$ if $i_1,\; i_2$ and $j_1,\; j_2$ belong to the different group respectively. Let
\bea
v &\to & v+M, \nonumber \\
x_i &\to & x_i+M \quad\mbox{for $i=1,\ldots ,r_1$},\nonumber \\
x_i &\to & x_i \quad\qquad\mbox{for $i=r_1+1 ,\ldots,N_1$},\nonumber \\
y_j &\to & y_j+M \quad\mbox{for $j=1,\ldots ,r_2$} ,\label{limit} \\
y_j &\to & y_j \quad\qquad\mbox{for $j=r_2+1,\ldots,N_2$} ,\nonumber \\
M &\to & \infty ,\nonumber\\
\eea 
while keeping
\be
\tLambda^{2 r_1 -r_2} = M^{-2k+2r_1 - r_2}\Lambda^{2k},
\ee
fixed. We obtain the limiting form of the  curve of (\ref{uspsu}):
\bea
\tLambda^{2r_1}M^{2k-2r_1+r_2}y^4 - M^{3k} y^3 &+& M^{4k-r_1}\prod_{i=1}^{r_1}(v-x_i) y^2 \nonumber \\ &&-\tLambda^{2r_1-r_2}M^{5k-2r_1}\prod_{j=1}^{r_2}(v-y_j) y + \tLambda^{3(2r_1-r_2)}M^{6k-6r_1+3r_2}=0. \nonumber
\eea
Making the shift $y \to M^{k-r_1}(-y)$ and throwing an overall factor away, the curve becomes 
\be
\tLambda^{2r_1-r_2}M^{-3r_1+r_2}y^4 +y^3 + \prod_{i=1}^{r_1}(v-x_i)y^2 + \tLambda^{2r_1-r_2}\prod_{j=1}^{r_2}(v-y_j)y + \tLambda^{3(2r_1-r_2)}M^{3(r_2-r_1)}=0. \label{ttt}
\ee
Notice that the term of $\tilde{y}^4$ always vanishes for $r_2 \le r_1 $ as $M \to \infty$. This result is in agreement with the result which is studied in \cite{dp}. When $r_2 < r_1$, as $M\to \infty$, this curve corresponds to $N=2\; SU(r_1)$ with $r_2$ flavor theory. When $r_1=r_2$ , the curve becomes that for $N=2\; SU(r_1)\times SU(r_2)$ with bifundamental matter theory.

\subsection{Solution in compact space and the symmetry}

For $USp$ case, we may easily present the infinite series expansion curve (\ref{uspsu}) as a polynomial by using (\ref{cm}), (\ref{hn}), (\ref{qh}). The polynomial $H(v)$ in (\ref{uspsu}) is
\be
H(v)=v^2\prod_{a=1}^{N_c}\left( v^2 - k_a^2\right) + A, \label{abc}
\ee
where
\be
A \equiv f(q)m^2 \prod_{a=1}^{N_c}(m^2-k_a^2).\label{A}
\ee
Taking care of the degree of the $H(v)$ for $v$ which is $2N_c +2$, we find the curve for softly broken $N=4\; USp(2N_c)$ as the polynomial,
\be
\sum_{n=0}^{2 N_c + 2}h_n(z)\frac{(-m)^n}{n!\beta^n} H^{(n)}(k)=0 ,\label{uspp}
\ee 
where
\bea
H^{(n)}(k) &= &\frac{\partial^n}{\partial k^n}H(k), \nonumber\\
k&=&v-\frac{m}{2}, \\
z &=&\frac{s}{\beta} , \\
\beta &=& -2\pi i.
\eea
Of course, since (\ref{uspp}) is equivalent to (\ref{usp}), this also satisfies all checks in the previous subsection. It is, however, very difficult to check such consistency in terms of (\ref{uspp}) for any $USp(2N_c)$.

Now we can read off the effect of the O4 plane in compact space. Since the $\theta_1(z|\tau)$ is an odd function of $z$, we have 
\be
h_n(-z)=(-)^n h_n(z).
\ee
Furthermore, since $H(k)$, (\ref{abc}), is an even function of $k$, we have
\be
H^{(n)}(-k) = (-)^n H^{(n)}(k).
\ee
Hence (\ref{uspp}) is invariant under 
\be
(s,v-\frac{m}{2}) \to (-s,-(v-\frac{m}{2})). \label{go4}
\ee
This appears to be the same symmetry as the O6 planes which extend along the directions the\\ $x^0,x^1,x^2,x^3,x^7,x^8,x^9$. Nevertheless, the interpretation is totally different. Let us consider (\ref{go4}) in a little more detail. To begin with, the O4 plane is produced by dividing the directions the $v,x^7,x^8,x^9$ by ${\bf Z_2}$. Other directions are the product space globally. This means that when $m=0$ since $X_m={\bf C}\times{\bf E}$ globally we may take normal O4 plane. As seen in (Fig \ref{covering2} (a)), this is trivial in the figure. In the curve level, when $m=0$ the $H(v-nm)$ becomes $H(v)$ which does not depend on $n$. Therefore (\ref{usp}) is merely $H(v)=0$. Accordingly, (\ref{go4}) means simply $v \to -v$. Now our space $X_m$ is ${\bf C}\times{\bf E}$ not globally but locally. We take the generalized O4 plane as ${\bf C}/{\bf Z_2}\times{\bf E}$ locally.\footnote{To be precisely, not ${\bf C}/{\bf Z_2}$ but $({\bf R}^3\times {\bf C})/{\bf Z_2}$ , where $x^7,x^8,x^9 \in {\bf R}^3$.} This operation is not, however, permitted globally except for $m=0$ in $X_m$ . Therefore there must be some effects for the ${\bf E}$ part globally. Actually the effect is seen as (\ref{go4}).

We give the explicit curve as the polynomial. For brevity, we use $k$ not $v$. For $N_c=1$, $H(k) = k^2(k^2 -u)+ A$, where $A =f(q)m^2(m^2-u)$. The curve (\ref{uspp}) becomes
\be
(k^4 -u k^2 + A) - h_1\tm (4 k^3 -2 u k) + h_2 \tm^2 (6 k^2 -u) - 4 h_3 \tm^3 k + h_4 \tm^4 =0 ,
\ee
where $\tm = \frac{m}{\beta}$. Let $k\to k+ \tm h_1$ to represent the curve in terms of $x,\; y$. Here $x$ is Weierstrass $\wp (z)$ function and $y = \frac{1}{2}\wp '(z)$. The above curve becomes
\bea
 k^4 &+& k^2(6\tm (h_2-h_1^2)-u) + 4\tm k(-2 h_1^3 + 3 h_1 h_2 - h_3)\nonumber \\
&+& \tm^4(-3 h_1^4 + 6h_1^2 h_2 - 4 h_1 h_3 + h_4) + \tm u (h_1^2 -h_2) + A=0.\label{nc1}
\eea
Use $x=-\partial_z \partial_z \: \ln \theta_1(z) - a_2(\tau )$, where $a_2$ is constant which does not depend on $z$. All combinations of $h_n$ in the elliptic Calogero Moser system are written in terms of $x,y$. 
\bea
h_1^2 - h_2 &=& x + a_2 ,\\
-2 h_1^3 + 3 h_1 h_2 - h_3 &=& 2y ,\\
-3 h_1^4 + 6h_1^2 h_2 - 4 h_1 h_3 + h_4&=& -3(x-a_2)^2 + \frac{1}{2}g_2 + 6a_2^2.
\eea
where $y^2 = (x-e_1(\tau))(x-e_2(\tau))(x-e_3(\tau))=x^3-\frac{1}{4}g_2(\tau) x-\frac{1}{4}g_3(\tau)$.
\\
Therefore, (\ref{nc1}) becomes\footnote{In \cite{uranga}, the same curve up to some constant factors is obtained by using O6 planes for $USp(2)$. Ref \cite{uranga} gives, in addition, equation (6.2) for $Sp(k)$ case. }
\bea
k^4 - k^2(6\tm^2 (x+a_2) + u) + 8\tm^3 y k &+& \tm^4 (-3(x-a_2)^2 +\frac{1}{2}g_2 + 6 a_2^2) \nonumber \\ &&+ \tm^2 u (x+a_2) + f(q)m^2(m^2-u)=0, \label{aa}
\eea
Since $USp(2)=SU(2)$, this curve must be equivalent to the one for $N=4 \quad SU(2)$. Let us confirm at least for $m\to 0,\infty$. First, for $\tm \to \infty$ , we already know that (\ref{aa}) becomes pure $N=2\; USp(2)$ curve by the argument in the previous subsection. It is known that the $N=2 \; USp(2)$ curve is equivalent to $N=2 \; SU(2)$ one \cite{as}. For $\tm \to 0$, throwing the overall factor away,  (\ref{aa}) becomes
\be
k^2-u=0
\ee
This is equivalent to the curve for $N=4\; SU(2)$ one with $m=0$.

For any $N_c$, we may construct the curve for softly broken $N=4\; USp(N_c)$ as polynomials except for an ambiguity of $f(q)$ by (\ref{uspp}). This ambiguity is fixed by comparing the discriminants of the $SU(2)$ and $USp(2)$ curves  since $f(q)$ is the same function for any $N_c$. The $USp(2)$ curve is, however, very complicated to look for the discriminant.

\subsection{$N=4 \; SO(N_c) $ theory and Decoupling limits}
\subsubsection{The $SO(2N_c)$ case}
In this subsection, we look for the softly broken $N=4 \; SO(2N_c)$ curve as the infinite series expansion by repeating the similar procedure. In this case the R-R charge of O4 plane is -1. Therefore, the effect of the O4 plane between NS5 branes appears as $v^{-2}$. Notice that there is no constant shift unlike the  $USp$ case. For example, the pure $N=2\; SO(2N_c)$ curve is 
\be
y^2 + v^{-2}\prod_{a=1}^{N_c}(v^2-k_a^2)y +1=0
\ee
To obtain the standard form, let $y\to v^{-2}y$. The curve becomes
\be
y^2 +\prod_{a=1}^{N_c}(v^2-k_a^2)y + v^4 =0.
\ee
For that reason, we find the curve for the softly broken $N=4 \; SO(2N_c)$,
\be
\sum_{n=-\infty}^{\infty}(-)^n q^{\frac{1}{2}n(n-1)}e^{n s}H(v-nm)=0, \label{so}
\ee
where
\be
H(v-nm)=(v-nm)^{-2}\prod_{a=1}^{N_c}((v-nm)^2 - k_a^2). \label{iii}
\ee
Here $k_a > 0$ are the classical positions of D4 branes and $-k_a<0$ for their mirrors. This curve satisfies (\ref{xmtf}),(\ref{O4}) and (\ref{go4}). We check the consistency of the curve (\ref{so}) by the diagram below.\\
\begin{tabular}{ccc}
$N=4 \quad SO(2N_c)$ & $\longrightarrow $ & $N=4 \quad SU(N_c)$ \\
 & (1) & \\
$\downarrow $(2) & & $\downarrow $ \\
&(3) & \\
$N=2\quad SO(2 N_1)$ with $N_2$ flavors & $\longrightarrow $& $N=2 \quad SU(r_1)$ with $r_2$ flavors\\
 &  &
\end{tabular}\\
As in the $SU$ and $USp$ cases, the flavor symmetry is gauged in (2) for some $N_1,N_2$.

For the limit (1), let
\bea
v_0 &\equiv &\frac{1}{N_c}\sum_{a=1}^{N_c}k_a\quad \to \quad \infty ,\\
v &\to & v + v_0 ,\\
k_a &\to & k_a + v_0 .
\eea
(\ref{iii}) becomes
\be
H(v-nm) \to v_0^{N_c-2}\prod_{a=1}^{N_c}(v-nm-k_a).
\ee
Here $v_0^{N_c-2}$ factor is thrown away since this is independent of $n$. We obtain the limiting form of the curve,
\be
\sum_{n=-\infty}^{\infty}(-)^n q^{\frac{1}{2}n(n-1)}e^{n s}\prod_{a=1}^{N_c}(v-nm - k_a)=0.
\ee
This is the curve for softly broken $N=4\; SU(N_c)$ as expected.

For the limit (2), let $N_1+N_2=N_c$ and 
\bea
k_a &\to & x_i \quad\qquad i=1,\ldots,N_1\mbox{ for }a=1,\ldots,N_1, \\
k_a &\to & y_j + m \quad j=1,\ldots,N_2\mbox{ for }a=N_1+1,\ldots,N_c ,\\
q & \to & 0,\\
m &\to & \infty,
\eea
while fixing
\bea
\Lambda^{b_0} &=& q m^{b_0}, \\
y&=&\omega m^{\frac{b_0}{2}}, \label{ppp}
\eea
where $b_0 = 4 N_1 -2N_2 -4$ , $\omega=e^s$. (\ref{ppp}) is required to keep at least one NS5 brane in the limiting configuration as in the $USp$ case.  We take $A(v)$ and $B(v)$ as
\bea
A(v)&\equiv & \prod_{i=1}^{N_1}(v^2-x_i^2), \\
B_{\pm }(v)&\equiv & \prod_{j=1}^{N_2}(v\pm y_j).
\eea
In this limit,
\bea
n=0\quad H&\sim &m^{2N_2} v^{-2}A(v), \\
n=\pm 1\quad H&\sim &m^{2N_1+N_2-2}B_{\pm}(v), \\
n\ne 0,\pm 1\quad H&\sim & m^{2(N_1+N_2)-2}.
\eea
The limiting form of the curve is 
\be
\frac{A(v)}{v^2} - y B_+(v) + y^2 \Lambda^{b_0}m^{-6N_1+4 N_2+6}-\frac{\Lambda^{b_0}}{y}B_-(v) + \frac{\Lambda^{3b_0}}{y^2}m^{-6N_1+4 N_2+6} = 0. \label{bbb}
\ee
When $-6N_1+4 N_2+6<0$, (\ref{bbb}) becomes
\be
B_+(v) y^2 - \frac{A(v)}{v^2}y  + \Lambda^{b_0} B_-(v) =0.
\ee
Let $v^2 B_+(v) y \to  y$, then we have the standard curve for $N=2\; SO(2N_1)$ with $ N_2$ flavors\cite{as,ha},
\be
y^2 - \prod_{i=1}^{N_1}(v^2-x_i^2)y + \Lambda^{b_0}v^4\prod_{j=1}^{N_2}(v^2- y_j^2)=0.
\ee
In particular, let us take $N_2=0$ , which satisfies the condition $-6N_1+4\cdot 0 +6 <0$. We obtain the pure $N=2\; SO(2N_1)$ curve.

 If $-6N_1+4 N_2+6=0$, $(N_1,N_2)=(2k+1,3k)$ $k=1,2,\ldots $. The curve (\ref{bbb}) becomes
\be
\Lambda^{b_0} y^4 - B_+(v) y^3 + \frac{A(v)}{v^2} y^2 - \Lambda^{b_0}B_-(v)y +\Lambda^{3b_0} =0, \label{ddd}
\ee
where $b_0 = 2k$. Since the flavor symmetry is gauged in this case, (\ref{ddd}) ought to represent the curve for $N=2\;SO(2N_1)\times SU(N_2)$ with bifundamental matters. To confirm this , let us check the limit (3).

As in the $USp$ case , we take the limits (\ref{limit}) where $r_2 \le r_1$, while keeping 
\be
\tLambda^{2r_1-r_2} =M^{-b_0 + 2r_1 - r_2}\Lambda^{b_0} .
\ee
fixed. Throwing an overall factor away and making the shift $y \to M^{k-r_1}(-y)$, the curve (\ref{ddd}) becomes
\be
\tLambda^{2r_1-r_2}M^{-3r_1+r_2}y^4 + y^3 + \prod_{i=1}^{r_1}(v-x_i) y^2 + \tLambda^{2r_1-r_2}\prod_{j=1}^{r_2}(v-y_j) y + \tLambda^{3(2r_1-r_2)}M^{3(r_2-r_1)} =0.
\ee
This is exactly the same as (\ref{ttt}). Therefore the same situation takes place. The first term always vanishes for $r_2\le r_1$ as $M\to \infty$. When $r_2 < r_1$ we have the curve for $N=2\;SU(r_1)$ with $r_2$ flavors and when $r_1=r_2$, the curve for $N=2\;SU(r_1)\times SU(r_2)$ with bifundamental matters. 

\subsubsection{The $SO(2N_c+1)$ case}

For the $SO(2N_c+1)$ case, the procedure is almost same as the $SO(2N_c)$ case.
The only difference is that we need one more D4 brane which is not dynamical and which is on the O4 plane. Therefore we multiply (\ref{iii}) by $(v-nm)$ to obtain
\be
H(v-nm)=(v-nm)^{-1} \prod_{a=1}^{N_c}((v-nm)^2 - k_a^2).
\ee
The curve for the $SO(2N_c+1)$ case is 
\be
\sum_{n=-\infty}^{\infty}(-)^n q^{\frac{1}{2}n(n-1)}e^{n s}(v-nm)^{-1} \prod_{a=1}^{N_c}((v-nm)^2 - k_a^2)=0.\label{so2}
\ee
As in the other cases, we can check the consistency of the curve (\ref{so2}) by the diagram below, \\
\begin{tabular}{ccc}
$N=4 \quad SO(2N_c+1)$ & $\longrightarrow $ & $N=4 \quad SU(N_c)$ \\
 & (1) & \\
$\downarrow $(2) & & $\downarrow $ \\
&(3) & \\
$N=2\quad SO(2 N_1+1)$ with $N_2$ flavors & $\longrightarrow $& $N=2 \quad SU(r_1)$ with $r_2$ flavors\\
 &  &
\end{tabular}\\
(\ref{so2}) satisfies also these checks. We do not put the details except for some comments on the limit (2), because the procedure is same as the other cases.

For the limit (2), $b_0=4N_1-2N_2-2$.  The condition for this limit is that $-6N_1+4N_2+3 \le 0$. When $-6N_1+4N_2+3=0$, the flavor group is gauged. However there is no solution for integers $N_1, N_2$. Therefore the flavor group is not gauged unlike the other cases.
 \\
 \\
{\it Conclusions for the $SO(N_c)$ case}

We find the curves for softly broken $N=4\; SO(N_c)$ as the infinite series expansion (\ref{so}) and (\ref{so2}) for $SO(2N_c)$ and $SO(2N_c+1)$, respectively. In this case, we have not recast these expansions into polynomials. Recall that $H(v)$ is a polynomial in obtaining (\ref{dhsu}) from (\ref{cm}).

\bigskip
\bigskip

%
%

\section*{Acknowledgments}
The author acknowledges Hiroshi Itoyama and Toshio Nakatsu for discussions on brane configurations and integrable systems. He is also grateful to Hiroshi Itoyama for a careful reading of the manuscript. He also thanks Asato Tsuchiya, Kazutoshi Ohta and Akira Tokura for useful discussions and comments.
\newpage

%
%

\end{document}